\documentclass{pasj00}
\draft

\begin{document}
\SetRunningHead{Tatematsu et al.}{N$_2$H$^+$ and HC$_3$N in Orion}
\Received{}
\Accepted{}

\title{N$_2$H$^+$ and HC$_3$N Observations of the Orion A Cloud}

\author{Ken'ichi \textsc{Tatematsu}, 
Ryo \textsc{Kandori},
Tomofumi \textsc{Umemoto},
and Yutaro \textsc{Sekimoto}}
\affil{National Astronomical Observatory of Japan, 
Mitaka, Tokyo 181-8588}
\email{k.tatematsu@nao.ac.jp, kandori@optik.mtk.nao.ac.jp, 
umemoto@hotaka.mtk.nao.ac.jp,  sekimoto.yutaro@nao.ac.jp}


%


\maketitle

\begin{abstract}
The ``$\int$-shaped filament'' of the Orion A giant
molecular cloud was mapped in N$_2$H$^+$ $J$ = 1$\rightarrow$0, 
and its northern
end, the OMC-2/3 region was observed also in HC$_3$N 
$J$ = 5$\rightarrow$4 and CCS $J_N$ = $4_3\rightarrow3_2$ and
$7_6\rightarrow6_5$.
The results are compared with maps of other molecular lines
and the dust continuum emission.
The N$_2$H$^+$ distribution is similar to the dust
continuum distribution,
except for the central part of the Orion Nebula.
The distribution of H$^{13}$CO$^+$ holds resemblance to that of 
dust continuum, but the N$_2$H$^+$ distribution looks more 
similar
to dust continuum distribution.
The N-bearing molecules, N$_2$H$^+$ and NH$_3$ seem to be more 
intense 
in OMC-2, 
compared with the H$^{13}$CO$^+$ and CS distribution.
This suggests that OMC-2 has higher abundance of 
N-bearing molecules or higher filling factor of the quiescent gas.
We identified 34 cloud cores from N$_2$H$^+$ data.
Their average physical parameters are 
$T_{ex}$ = 9.2$\pm$4.2 K, 
$\Delta v$ = 0.92$\pm$0.52 km s$^{-1}$, 
$R$ = 0.086$\pm$0.025 pc, and
$M$ = 46$\pm$32 $M_{\odot}$.
The masses of cores identified in both N$_2$H$^+$ and
H$^{13}$CO$^+$
in the OMC-2/3 region are rather consistent. 
Over the Orion Nebula region, the N$_2$H$^+$ linewidth is large 
(1.1$-$2.1 km s$^{-1}$).
In the OMC-2/3 region, it becomes moderate (0.5$-$1.3 km s$^{-1}$), and
it is smaller (0.3$-$1.1 km s$^{-1}$) in the south of the Orion Nebula.
On the other hand, the gas kinetic temperature of the quiescent cores 
observed in 
N$_2$H$^+$ is rather constant ($\sim$ 20 K)
over the $\int$-shaped filament.
The average N$_2$H$^+$ excitation temperature in Orion cores is 
$\sim$ 1.6 times as high as that in Taurus cores ($\sim$ 5.7 K).
The N$_2$H$^+$ excitation temperature 
decreases toward the south, suggesting
the core gas density or N$_2$H$^+$ abundance 
decreases toward the south.
We found one peculiar H$^{13}$CO$^+$ core which is not prominent 
in either
N$_2$H$^+$, HC$_3$N, or dust.
This core overlaps with the lobe of the intense outflow from
a nearby protostar.
We detected no CCS emission in the OMC-2/3 region.
In general, N$_2$H$^+$ and HC$_3$N distribution is quite similar 
in the OMC-2/3 region,
but we observed displacement between N$_2$H$^+$ and HC$_3$N 
over 2$\arcmin$ scale in OMC-3, which
has a chain of Class 0-I protostars (candidates).
This displacement might be due to either chemical evolution or effect of 
protostellar 
outflows.

\end{abstract}

\section{Introduction}

Most of stars in the Galaxy form in giant molecular clouds.
The molecular cloud core is known to be the site of star formation
(e.g., \citet{bei86}).
The evolution of molecular cloud cores 
in giant molecular clouds is less known,
compared with that of low-mass dark cloud cores.
In dark clouds, molecules such as CCS, HC$_3$N, NH$_3$, and N$_2$H$^+$,
and the neutral carbon atom C$^0$
are known to be good tracer of chemical evolution
(e.g., \cite{hir92,suz92,mae99,hir02}).
The carbon chain molecules, CCS and HC$_3$N tend to trace 
the early chemical 
evolutionary stage, whereas NH$_3$ and N$_2$H$^+$ tend to trace
the later stage.
It is now well known that N$_2$H$^+$, which is less affected by 
depletion, is one of the best molecular
tracers for low-mass star forming regions 
(e.g., \cite{cas99,aik01,ber01,ber02,cab02,zli02,taf02,lee03,she03}).

The Orion A cloud is an archetypal 
giant molecular cloud.
This cloud has been extensively mapped in $^{12}$CO $J$ = 1$\rightarrow$0
\citep{hey92}, 
in $^{13}$CO $J$ = 1$\rightarrow$0 \citep{bal87,nag98}, 
in CS $J$ = 1$\rightarrow$0 \citep{tat93},
in CS $J$ = 2$\rightarrow$1 \citep{tat98}, 
in NH$_3$ \citep{bat83,ces94}, 
and in H$^{13}$CO$^+$ $J$ = 1$\rightarrow$0 \citep{ike07}.
The dust continuum emission in the Orion A cloud was studied by
\citet{chi97}, \citet{lis98}, and \citet{joh99}.
The region near Orion KL was studied in many molecular lines
by \citet{ung97}.

Of particular interest is the OMC-2/3 region, which is the 
northernmost part of the Orion A cloud. This region
is one of the
best studied intermediate-mass star forming regions.
\citet{chi97} investigated the distribution of the dust
continuum emission and found at least 21 protostellar dust condensations.
\citet{aso00} observed H$^{13}$CO$^+$ and CO in this region, and
identified cloud cores and outflows.
Outflows were also observed by \citet{yu00,wil03}.
\citet{tak06} studied a molecular cloud core and an outflow associated
with the protostar MMS7 in detail.
\citet{tsu02,tsu03} investigated the properties of the X-ray sources 
observed with the
Chandra Observatory, and identified/classified them on the basis of 
near infrared observations (Class I, Class II, and Class III+MS, 
brown dwarf candidates).
\citet{tsu01} detected the X-ray emission from two protostar
candidates in OMC-3. 
\citet{tsu04} concluded that one of these two X-ray sources
is the
shock-induced X-ray emission by protostellar jet from 
Class I protostar(s).
\citet{joh03} observed seven submillimeter sources in eight 
molecular lines
and investigated the physical condition.

In this study, we revisit the Orion A cloud, including the OMC-2/3 region, 
through new molecular-line
observations.
The purposes of this study are 
(1) to make clear
the distribution of the quiescent molecular gas and to compare it
with star formation activity, 
(2) to describe the chemical evolution in the OMC-2/3 region, 
(3) to compare the physical properties of N$_2$H$^+$ 
cores in the Orion massive star-forming region
with those in the low-mass star forming dark cloud found in Taurus,
and (4) to investigate variation in N$_2$H$^+$ core properties
along the ``$\int$-shaped filament.''

The distance to the Orion A cloud is assumed to be 450 pc 
\citep{gen89}. At this distance,
1$\arcmin$ corresponds to 0.13 pc.

\newpage

\section{Observations}

Observations were carried out by using the 45 m radio telescope
of Nobeyama Radio Observatory from 2005 May 11 to 20,
and from 2007 March 4 to 9.  
The employed receiver front ends were
the 25-element focal-plane SIS array receiver ``BEARS'', and
the single-element SIS receivers, ``S100'' and ``S40''.
We observed 
N$_2$H$^+$ $J = 1\rightarrow0$ at 93.1737767 GHz \citep{cas95}
by using receiver ``BEARS'',
CCS $J_N$ = 4$_3$-3$_2$ at 45.379033 GHz \citep{yam90}
by using receiver ``S40'', 
CCS $J_N$ = 7$_6$-6$_5$ at 81.505208 GHz \citep{hir06}
by using receiver ``S100'', 
and 
HC$_3$N $J$ = 5-4 at 45.490316 GHz \citep{yam90}
by using receiver ``S40''.
CCS and HC$_3$N were observed simultaneously, by using 
two receivers with the polarization
splitter.
The half-power beamwidth of the element beam of
BEARS was 17$\farcs$8$\pm$0$\farcs$4 at 93 GHz.
Those of S40 and S100 are 38$\farcs$4$\pm$0$\farcs$1 
and 18$\farcs$2$\pm$0$\farcs$1 at 43 and 86 GHz, respectively.
The beam separation of ``BEARS'' is 41$\farcs$1. 
The spacing grid employed in mapping observations with ``BEARS'' was 
20$\farcs$55, which is half of the beam separation of 
elements within the array and close to the
half-power beamwidth.
The spacing grid employed in ``S40'' and ``S100''observations was 
40$\arcsec$.
The receiver back end for ``BEARS'' was a digital autocorrelator
and that for  ``S40'' and ``S100'' was acousto-optical spectrometers.
The spectral resolution for ``BEARS'' was 37.8 kHz 
(corresponding to $\sim$
0.12 km s$^{-1}$ at 93 GHz) and that for  ``S40'' and ``S100''
was 37 kHz  (corresponding to $\sim$ 0.25 km s$^{-1}$ at 45 GHz and
0.14 km s$^{-1}$ at 81 GHz).
The map reference center is Orion KL
at 
R. A. (J2000) = 5$^h$ 35$^m$ 14$^s$.5, 
Dec. (J2000) = $-$5$\degree$ 22$\arcmin$ 30$\arcsec$.
Spectra were obtained in the position-switching mode.
The employed off position is either ($\Delta$ R. A., $\Delta$ Dec.)
= ($-$30$\arcmin$, 5$\arcmin$) or (0$\arcmin$, $-$100$\arcmin$)
with respect to Orion KL.
We calibrated the gain of each ``BEARS'' element and
obtained the absolute intensity scale
in the way described in \citet{tat04}.
The intensity is reported in terms of the corrected
antenna temperature $T_A^*$.
The main-beam efficiency 
is 0.77$\pm$0.03 for ``S40'', 
and  0.51$\pm$0.02 for ``S100'' and ``BEARS''.
The telescope pointing was established by observing Orion KL
in the 43-GHz SiO maser line every 1$-$1.5 hours.
The data were reduced by using the software package NewStar
of Nobeyama Radio Observatory and IDL of Research Systems, Inc.

\section{Results and Discussion}

\subsection{N$_2$H$^+$ Observation} 

Figures \ref{fig:figure1} and \ref{fig:figure2} show the distribution 
of the velocity-integrated intensity of
the N$_2$H$^+$ $J = 1\rightarrow0$ 
$F_1$ = 2$\rightarrow$1 hyperfine group. 
Because the linewidth is broad, we cannot separate hyperfine 
components well.
Therefore, we integrated the three components including 
the most intense main component.
The total length of the filament is about 12 pc, and
the figures exhibit filamentary and clumpy structure well.

\begin{figure}
  \begin{center}
  \end{center}
  \caption{The grey-scale map of the velocity-integrated intensity 
of the N$_2$H$^+$ 
$J = 1\rightarrow0$ 
$F_1$ = 2$\rightarrow$1 hyperfine group.  The level interval for 
the contour is 0.749 K km s$^{-1}$.
Pluses represent Orion KL (labeled as ``Orion KL'') and cores identified in
N$_2$H$^+$.}\label{fig:figure1}
\end{figure}

\begin{figure}
  \begin{center}
  \end{center}
  \caption{The coutour map of the velocity-integrated intensity 
of the N$_2$H$^+$ 
$J = 1\rightarrow0$ 
$F_1$ = 2$\rightarrow$1 hyperfine group.  The contour levels 
are 0.749 K km s$^{-1} \times$ (1, 2, 4, 8).
The intensity maxima of the N$_2$H$^+$ cores are shown as
pluses.}\label{fig:figure2}
\end{figure}

Figures \ref{fig:figure3} and 
\ref{fig:figure4} show a comparison of N$_2$H$^+$ $J = 1\rightarrow0$ 
$F_1$ = 2$\rightarrow$1 hyperfine group map with 850 $\mu$m thermal 
dust continuum emission map
\citep{joh99} in a closer view of Orion KL.
These maps show that the N$_2$H$^+$ distribution
resembles the dust continuum distribution very well.
For example, V-shaped structure about 15$\arcmin$ south of
Orion KL is very clear in both maps.
However, it should be noted that the Orion bar,
which is prominent in the dust continuum, is not seen
in N$_2$H$^+$.
N$_2$H$^+$ is known to be sensitive to quiescent molecular gas,
while the dust continuum is sensitive to the photodissociation
region (PDR).
The dust distribution is thinner in width and more compact 
than the N$_2$H$^+$ distribution,
because structures with scales larger than the largest chopper
throw (65$\arcsec$) are missing from the dust continuum map.

\begin{figure}
  \begin{center}
  \end{center}
  \caption{The velocity-integrated map of the N$_2$H$^+$ 
$J = 1\rightarrow0$ 
$F_1$ = 2$\rightarrow$1 hyperfine group in a closer view of Orion KL.  
The level interval for 
the contour is 0.749 K km s$^{-1}$. 
Pluses represent Orion KL (labeled as ``Orion KL'') and cores identified in
N$_2$H$^+$.}\label{fig:figure3}
\end{figure}

\begin{figure}
  \begin{center}
  \end{center}
  \caption{The 850 $\mu$m thermal dust continuum emission map
\citep{joh99}.  This
map covers the same Dec. range of Figure \ref{fig:figure2}.  The
image is convolved for the angular resolution of the
N$_2$H$^+$ observation.  
The level interval for the contour is 0.84 Jy/beam.
Pluses represent cores identified in
N$_2$H$^+$.}\label{fig:figure4}
\end{figure}

We identified N$_2$H$^+$ cores by using two sets of the velocity 
channel maps.
The first set is 0.3 km s$^{-1}$-width velocity channel maps
of the $F_1$, $F$ = 0, 1$\rightarrow$1, 2 component,
which is an isolated component of the seven hyperfine components.
The advantage of this set is that we can separate molecular gas having
different velocities easily.  
However, in some cases the signal-to-noise 
ratio
is not sufficient.
The second set is  0.5 km s$^{-1}$-width velocity channel maps
of $F_1$ = 2$\rightarrow$1 hyperfine group, which contains
the most intense hyperfine component
$F_1, F$ = 2, 3$\rightarrow$1, 2
and two neighboring hyperfine components.
This hyperfine group has three components in the
velocity range of less than 2 km s$^{-1}$,
and it is hard to derive detailed velocity structure.
On the other hand, we can detect molecular gas having weaker
intensities with this set.
By taking advantages of these two sets, we have
identified a total of 34 molecular cloud cores in N$_2$H$^+$.
Table 1 summarizes the identification by eye and properties of 
N$_2$H$^+$ cores.
The value after $\pm$ in the antenna temperature,
excitation temperature, optical depth, and linewidth
shows the 1$\sigma$ uncertainty in the hyperfine fitting to
the spectrum toward the intensity maximum.
The bottom row with "ave" lists the average and standard deviation
for all cores.

The positions of the intensity maxima of the identified
cores are illustrated in Figures \ref{fig:figure1}, \ref{fig:figure2},
and \ref{fig:figure3}.
Orion KL is not prominent in N$_2$H$^+$, and is not cataloged
as a core. It is located on the eastern bay of the N$_2$H$^+$ 
emission ridge.
The basic physical parameters of the identified cores 
are summarized in Tables 1 and 2.
The HWHM (half of FWHM) core radius $R$ is measured as $\sqrt{S}/\pi$ 
($S$ is the core area
S at the half maximum), and then corrected for the
telescope beam size. 
We fit the spectrum observed toward the intensity maximum 
by using the hyperfine spectrum model
consisting of multiple Gaussian components including the
effects of optical depth assuming a single excitation temperature.
Figures \ref{fig:figure5}, \ref{fig:figure6}, \ref{fig:figure7}, 
\ref{fig:figure8}, and \ref{fig:figure9} show the examples of the 
hyperfine fitting.
The procedure is given in \citet{tat04}.
The intrinsic relative intensities of the hyperfine components
are taken from \citet{tin00}.
The free parameters are the excitation temperature $T_{ex}$,
the sum of optical depths of the hyperfine components 
$\tau_{TOT}$, systemic
velocity (radial velocity), and intrinsic linewidth (which is
corrected for broadening due to line optical depth and instrumental
resolution).
$T_A^*$ in the table represents the intensity of the main N$_2$H$^+$ 
$J = 1\rightarrow0$
component ($F_1$, $F$ = 2, 3$\rightarrow$1, 2) 
derived in the hyperfine fitting. 
When the fitting is not very successful,
we show approximate $T_A^*$ values by indicating with ``$\sim$''.
$\tau_{TOT}$ represents the total optical depth summing up those of
the seven hyperfine components of the N$_2$H$^+$ emission.  
The
optical depth of the main N$_2$H$^+$ $J = 1\rightarrow0$
component ($F_1$, $F$ = 2, 3$\rightarrow$1, 2) is 0.259 $\times$ 
$\tau_{TOT}$
\citep{tin00}.
The details of the column density estimation are given in
\citet{cac02}.
The H$_2$ column density $N$(H$_2$) is derived by assuming 
the N$_2$H$^+$ fractional abundance
relative to H$_2$ is 3.0 $\times$ 10$^{-10}$ \citep{cas02}.
The average density $n$(H$_2$) is derived as $N$(H$_2$)/ 2 $R$.
For the OMC-2/3 region, we include
the existence of molecular outflows and
near-infrared H$_2$ jets (\cite{aso00} and  references therein)
as ``O'' and ``H'', respectively, in Table 2.

The N$_2$H$^+$ excitation temperature toward the core intensity peak is
$T_{ex} = 9.2\pm4.2$ K (average and standard deviation). 
This value is 1.6 times as high as
that in Taurus cores (
$T_{ex} = 5.7\pm1.2$ K for the
central nine positions of each core in the
Taurus cloud, 
which is good for
comparison with Orion cores by taking into account that Orion is 
three times more distant, \cite{tat04}).
The optical depth of the main component 
($F_1$, $F$ = 2, 3$\rightarrow$1, 2)
is found to be moderate (1.1$\pm$0.7) for the intensity peak,
and this is similar to the situation in Taurus cores \citep{tat04}.

\begin{figure}
  \begin{center}
  \end{center}
  \caption{The N$_2$H$^+$ spectrum toward the 
intensity maximum position
of N$_2$H$^+$ core 4.  The best-fit hyperfine fitting result is shown
as a smooth curve.  The velocity axis is for 
the main N$_2$H$^+$ $J = 1\rightarrow0$
component ($F_1$, $F$ = 2, 3$\rightarrow$1, 2).}\label{fig:figure5}
\end{figure}

\begin{figure}
  \begin{center}
  \end{center}
  \caption{The same as Figure \ref{fig:figure5} but for N$_2$H$^+$ 
core 6}\label{fig:figure6}

\end{figure}

\begin{figure}
  \begin{center}
  \end{center}
  \caption{The same as Figure \ref{fig:figure5} but for N$_2$H$^+$ 
core 23}\label{fig:figure7}
\end{figure}

\begin{figure}
  \begin{center}
  \end{center}
  \caption{The same as Figure \ref{fig:figure5} but for N$_2$H$^+$ 
core 27}\label{fig:figure8}
\end{figure}

\begin{figure}
  \begin{center}
  \end{center}
  \caption{The same as Figure \ref{fig:figure5} but for N$_2$H$^+$ 
core 32}\label{fig:figure9}
\end{figure}

We summarize the average physical parameters of N$_2$H$^+$ cores
with standard deviations.
The line width is $\Delta v$ = 0.92$\pm$0.52 km s$^{-1}$,
core size is $R$ = 0.086$\pm$0.025 pc,
column density is $N$(H$_2$) = (6.7$\pm$2.6)$\times$10$^{22}$ cm$^{-2}$,
average density is $n$(H$_2$) = (1.2$\pm$0.4)$\times$10$^5$ cm$^{-3}$,
and core mass is $M$ = 46$\pm$32 $M_{\odot}$.
We derived the virial parameter, which
is defined as the virial mass
divided by the core mass,
to be 0.39$\pm$0.32.
One possibility is that the actual N$_2$H$^+$ abundance is 
higher than the value
we assumed.
However, the mass estimation usually accompanies
uncertainty by a factor of 2-3.
Furthermore, it seems that the N$_2$H$^+$ abundance
varies over the $\int$-shaped filament (see discussion later).
Therefore, it would be hard to derive a precise value
of N$_2$H$^+$ abundance on the basis of 
simple analysis assuming virial equilibrium.

When we compare Orion N$_2$H$^+$ cores with Taurus N$_2$H$^+$ cores
\citep{tat04},
Orion cores have 2.6, 3.2, 4.0 and 32 times larger core size,
linewidth, H$_2$ column density, and core mass than
Taurus cores, respectively.
The H$_2$ column density in Taurus used here is an average for 
central nine positions taking into account the difference in distance.

Figures \ref{fig:figure10} and \ref{fig:figure11} 
show a comparison between N$_2$H$^+$ (present study) 
and H$^{13}$CO$^+$ \citep{aso00}
in a closer view of the OMC-2/3 region.
Basically, the distribution of these tracers is similar.
Close inspection lets us see some differences.
First, the H$^{13}$ CO$^+$ core AC9 is not prominent 
in N$_2$H$^+$.
Second, N$_2$H$^+$ core 4
does not have a tail toward the south, which is seen in the H$^{13}$ 
CO$^+$
and in the dust continuum (Figure \ref{fig:figure4}).
Third, N$_2$H$^+$ cores 10 and 12 are more prominent in 
N$_2$H$^+$ than
H$^{13}$ CO$^+$ counterparts.

\begin{figure}
  \begin{center}
  \end{center}
  \caption{The velocity-integrated map of the N$_2$H$^+$ 
$J = 1\rightarrow0$ 
$F_1$ = 2$\rightarrow$1 hyperfine group in a closer view of the OMC-2/3 region.  
The level interval for 
the contour is 0.749 K km s$^{-1}$.
Pluses represent cores identified in
N$_2$H$^+$.
Thin straight lines represent the border between OMC-2
and OMC-3 used by \citet{chi97}}\label{fig:figure10}
\end{figure}

\begin{figure}
  \begin{center}
  \end{center}
  \caption{Reproduction of H$^{13}$CO$^+$ map of \citet{aso00}, 
but in
J2000 coordinates. This
map covers the same R. A. and Dec. range of 
Figure \ref{fig:figure10}. Contour intervals are 0.36 K km s$^{-1}$ 
starting at 0.36 K km s$^{-1}$. The
stars mark the 1.3 mm continuum sources \citep{chi97}, and the
triangles mark the 350 km continuum sources 
\citep{lis98}.}\label{fig:figure11}
\end{figure}

Next, we investigate the variation of the core properties along the
``$\int$-shaped filament.''
Figure \ref{fig:figure12} plots the N$_2$H$^+$ linewidth 
against the declination.
By using the intrinsic N$_2$H$^+$ linewidth,
we investigate the velocity dispersion of
the quiescent gas along the ``$\int$-shaped filament.''
Over the Orion Nebula region, the linewidth is large 
(1.1$-$2.1 km s$^{-1}$).
In the OMC-2/3 region, it becomes moderate (0.5$-$1.3 km s$^{-1}$), and
it is smaller (0.3$-$1.1 km s$^{-1}$) in the south of the
Orion Nebula.
It is important that we observe the tendency in N$_2$H$^+$, which
is the molecule least affected from star formation activities 
such as outflows
\citep{wom93}.
It was suggeted that
larger linewidths of the quiescent gas to form stars
will lead to higher mass accretion rate onto protostars
and eventually to more massive stars
(see, e.g., \cite{tat93}).
An argument against this idea is that larger linewidths might be
a result of more massive star formation.
The present study shows a trend in linewidth of the quiescent gas, and
might suggest that
variation in linewidth could serve as difference in the initial condition
for on-going star formation in the Orion cloud.
The core radius does not show any prominent trend against the declination
(figure not shown).
Figure \ref{fig:figure13} shows the temperature variation along
the ``$\int$-shaped filament.''
The kinetic temperature from CO $J$ = 3$\rightarrow$2 seems
to trace the outer, warmer, less-dense layer of the cloud
externally heated (\cite{wil99}; see also \citet{cas90}).
The NH$_3$ rotation temperature is about 20 K over the 
filament, implying
that the gas temperature of the quiescent dense gas is rather constant.
The average and standard deviation of the NH$_3$ rotation temperature
are 21.6 and 5.3 K, respectively.
A rotation temperature of $\sim$ 20 K corresponds to 
a gas kinetic temperature
of $\sim$ 20 K \citep{dan88}.
Figure 3 of \citet{wil99} shows a clear variation in rotation
temperature.  However, at positions coincident with quiescent
cloud cores identified in N$_2$H$^+$ and CS,
this variation is not seen.
It is very interesting that the quiescent Orion cores show 
constant kinetic temperatures while their linewidth varies
along the filament.
The kinetic temperature in quiescent Orion cores are
twice as high as that in dark cloud cores
(8-10 K, see, e.g., \citet{ben80,tat99})
The N$_2$H$^+$ excitation temperature decreases toward the 
south slightly,
which means that the core density or N$_2$H$^+$ abundance 
decreases toward the south.
\citet{tat93} and \citet{tat98} suggested that the Orion A cloud has
the average density decrease toward the south.
In detail, the excitation temperature seems to have its peak
around OMC-2.

\begin{figure}
  \begin{center}
  \end{center}
  \caption{The N$_2$H$^+$ linewidth 
against the declination.
Labels ``OMC-2'' and ``OMC-3'' represent the positions
of N$_2$H$^+$ cores 4 and 10, respectively.}\label{fig:figure12}
\end{figure}

\begin{figure}
  \begin{center}
  \end{center}
  \caption{The temperature 
against the declination.
The gas kinetic temperature from CO $J$ = 3$\rightarrow$2 
and
NH$_3$ rotation temperature \citep{wil99} of cores identified in both
CS and N$_2$H$^+$, and
N$_2$H$^+$ excitation temperature (present study) of N$_2$H$^+$ cores are 
shown.  The solid horizontal line represents 20 K.}\label{fig:figure13}
\end{figure}

H$^{13}$CO$^+$ in the Orion A cloud was observed by \citet{aso00} 
and \citet{ike07}.  The distribution in these maps is consistent
in the OMC-2/3 region, which was covered by both studies.
\citet{ike07} used automated software to identify cores, while
\citet{aso00} identified cores through visual inspection.
The former identified two times as many cores as \citet{aso00} 
in the OMC-2/3 region.
Part of this difference might be due to different sensitivities,
but we suspect that different core identification methods are 
the main reason
for disparate core numbers.
For comparison, we use \citet{aso00}, which identified cores
through visual inspection, for consistency.
The cross identification is given in Table 2.
The column ``Aso'' lists H$^{13}$ CO$^+$ cores in \citet{aso00},
and column ``Paper I'' lists CS $J$ = 1$\rightarrow$0 cores in 
\citet{tat93}.

Using the cores identified in both N$_2$H$^+$ and H$^{13}$CO$^+$,
we compare the N$_2$H$^+$ and H$^{13}$CO$^+$ cores
in the OMC-2/3 region 
(Figures \ref{fig:figure14}, \ref{fig:figure15}, and 
\ref{fig:figure16}).
The N$_2$H$^+$ linewidth tends to narrower than the H$^{13}$CO$^+$
linewidth, which is consistent with the fact that N$_2$H$^+$
traces the quiescent molecular gas.
The core radius and core mass are rather consistent between these 
two lines,
and it is likely that these two molecular lines trace similar 
density regions, although the linewidth is different to some extent.
On the other hand, only eight cores out of 14 N$_2$H$^+$ (cores 1 
through
14) in the OMC-2/3 region
have H$^{13}$CO$^+$ counterparts.

\begin{figure}
  \begin{center}
  \end{center}
  \caption{The N$_2$H$^+$ linewidth 
and H$^{13}$CO$^+$ linewidth in the OMC-2/3 region.  Starless cores
and star-forming cores are based on classification in 
\citet{aso00}.}\label{fig:figure14}
\end{figure}

\begin{figure}
  \begin{center}
  \end{center}
  \caption{The N$_2$H$^+$ core radius 
and H$^{13}$CO$^+$ core radius in the OMC-2/3 region}\label{fig:figure15}
\end{figure}

\begin{figure}
  \begin{center}
  \end{center}
  \caption{The N$_2$H$^+$ core mass 
and H$^{13}$CO$^+$ core mass in the OMC-2/3 region}\label{fig:figure16}
\end{figure}

N$_2$H$^+$ cores 10 and 12 are more prominent in N$_2$H$^+$ than
H$^{13}$ CO$^+$ counterparts in OMC-2. 
\cite{tat93} showed that,
in
a region containing Orion KL and OMC-2 (see their Figure 4),
NH$_3$ tends to be stronger in the north (OMC-2) while
CS tends to be stronger in the south.
\citet{ung97} studied molecular distribution around Orion KL
(from 6$\arcmin$ south to 6$\arcmin$ north),
and found that N$_2$H$^+$ tends to be stronger in the north,
H$^{13}$ CO$^+$ shows similar trend but less prominent,
and other molecules show different trends
(peaked around Orion KL or flat).
Figure  \ref{fig:figure17} shows the antenna temperature of 
N$_2$H$^+$ cores (present study) 
and H$^{13}$CO$^+$ cores \citep{ike07} 
against the declination. 
The dashed and solid lines delineate the rough 
upper boundary
of the antenna temperature of N$_2$H$^+$
and H$^{13}$CO$^+$, respectively, by connecting
local maxima to guide the reader's eye for the global trend. 
N$_2$H$^+$ is intense in OMC-2 (N$_2$H$^+$ cores 10 and 12) 
with respect to the H$^{13}$CO$^+$ intensity variation.
Figure \ref{fig:figure18} shows the antenna temperature of 
N$_2$H$^+$ cores (present study) 
and CS cores \citep{tat93} 
against the declination. 
Again, N$_2$H$^+$ is relatively intense 
in OMC-2 (N$_2$H$^+$ cores 10 and 12) with respect 
to the CS
intensity variation.
The N-bearing molecules, N$_2$H$^+$ and NH$_3$ seem to be more intense 
in OMC-2 (N$_2$H$^+$ cores 10 and 12), 
compared with the H$^{13}$CO$^+$ and CS distribution.
The gas kinetic temperature of quiescent cores 
is rather constant judging from the
NH$_3$ rotation temperature.
This suggests that OMC-2 has 
higher abundance of 
N-bearing molecules or higher filling factor of the quiescent gas.
Although OMC-2 is known as active star cluster forming region
\citep{joh90},
it seems that it still has a large reservoir of quiescent gas
available for
future star formation.
Note that N-bearing molecules, N$_2$H$^+$ and NH$_3$ are 
late-type molecules.
\citet{chi97} showed that OMC-3 has
Class 0 protostars (candidates) and suggested 
evolutionary trend from north (OMC-3, younger) 
to south (OMC-2, older).
This trend is consistent with 
the trend in the chemical properties (N-bearing late-type molecules
are more abundant in OMC-2).

\begin{figure}
  \begin{center}
  \end{center}
  \caption{The antenna temperatures of N$_2$H$^+$ (present study) 
and H$^{13}$CO$^+$ \citep{ike07} 
against the declination.
The dashed and solid lines connect local maximum intensities
in N$_2$H$^+$ and H$^{13}$CO$^+$, respectively, to
guide the reader's eye for the global trend.
}\label{fig:figure17}
\end{figure}

\begin{figure}
  \begin{center}
  \end{center}
  \caption{The antenna temperature of N$_2$H$^+$ (present study) 
and CS \citep{tat93} 
against the declination.
The dashed and solid lines connect local maximum intensities
in N$_2$H$^+$ and CS, respectively, to
guide the reader's eye for the global trend.}\label{fig:figure18}
\end{figure}

\subsection{HC$_3$N and CCS Observations} 

Figure \ref{fig:figure19} compares the N$_2$H$^+$ with HC$_3$N 
distribution 
in the OMC-2/3.
Globally, their distribution is quite similar,
although they are known to be late-type and early-type molecules 
in dark cloud
chemistry, respectively \citep{hir92}.
Both are intense in OMC-2.
This is in contrast with
the fact that OMC-2 is not very prominent compared with 
OMC-3
in H$^{13}$CO$^+$.
OMC-2 is known as a
cluster
forming region \citep{joh90}.
Although HC$_3$N is known to be a tracer of early chemical 
evolutionary stage
in dark clouds \citep{hir92}, 
it is not always so.
\citet{ung97} shows that the HC$_3$N $J$ = 10$\rightarrow$9 and 
12$\rightarrow$11 emission are peaked at 
Orion KL very clearly,
although it shows prominent star formation already.
On the other hand, Figure \ref{fig:figure4} shows that 
N$_2$H$^+$ is rather
weak near Orion KL, suggesting that very high temperature 
due to star formation
activity destroys this molecule due to CO evaporation 
from the dust.
The HC$_3$N linewidth around N$_2$H$^+$ cores 9 and 10
in OMC-2 is $\sim$ 1.2 km s$^{-1}$ and similar to
or only slightly larger than the N$_2$H$^+$ linewidth.
HC$_3$N seems to be abundant in dense, quiescent gas, even
with prominent star formation activities.

\begin{figure}
  \begin{center}
  \end{center}
  \caption{The N$_2$H$^+$ distribution 
with the HC$_3$N distribution
in the OMC-2/3 region. The grey scale represents N$_2$H$^+$ and
contours represent HC$_3$N.
The contour interval is 0.227 K km s$^{-1}$.
The grey scale ranges from 0.2 to 8.0 K km s$^{-1}$, 
while the maximum intensity of the N$_2$H$^+$ emission 
is 9.3 K km s$^{-1}$.
The crosses represent submillimeter protostars (candidates) by
\citet{chi97}}\label{fig:figure19}
\end{figure}

We found the displacement between N$_2$H$^+$ and HC$_3$N in
OMC-3 (Figure \ref{fig:figure20}).
HC$_3$N shows two emission peaks on both sides of N$_2$H$^+$
core 4.
N$_2$H$^+$
core 4 is associated with two X-ray protostar candidates
TKH8 and TKH10
\citep{tsu01} (Figure \ref{fig:figure21}).
Figure \ref{fig:figure22} shows HC$_3$N stamp map (profile map) near
TKH8 and TKH10.
\citet{tsu04} concluded that one of 
the X-ray emitting protostars,
TKH8, is a Class I protostar emitting the X-ray
emission from the protostellar jet.
A possibility is that
the displacement represents the chemical evolution.
Another possibility is that the degree of molecular
depletion varies along the ridge
of OMC-3.
It is interesting to check whether there is temperature
variation.
\citet{wil99} shows that CS cores 3, 4, and 5 of \citet{tat93}
located in OMC-3
have NH$_3$ rotation temperature
$T_{rot}$ = 15, 19, and 20 K, respectively.
CS core 3 is located at R. A. (J2000) = 5$^h$ 35$^m$ 17$^s$.6, 
Dec. (J2000) = $-$5$\degree$ 0$\arcmin$ 30$\arcsec$, and 
corresponds to a HC$_3$N peak on NW side of N$_2$H$^+$ core 4.
N$_2$H$^+$ cores 4, 5, and 6 have an N$_2$H$^+$ excitation 
temperature of
$T_{ex}$ = 9.3, 7.5, and 6.6 K, respectively.
The dust spectral index is rather constant in the OMC-2/3 
region except 
for MMS6
\citep{lis98,joh99}, and the OMC-2/3 region has no
dust temperature variation except for MMS6.
Because there is no evidence that N$_2$H$^+$ core 4 has lower
temperature in OMC-3, 
it is not easy to explain the displacement
between HC$_3$N and N$_2$H$^+$ in terms of different degrees of
depletion.
\citet{tsu04} showed that the Class I X-ray protostar TKH8
has near-infrared H$_2$ jets toward the west, although this
jet is very small in size.
\citet{aso00} showed larger-scale CO outflows in this region.
N$_2$H$^+$ is known to trace the quiescent gas, and
will be suppressed in outflow affected gas.
The displacement between HC$_3$N and N$_2$H$^+$ peaks
observed in OMC-3 could represent
the chemical evolution or the effect of molecular outflow
from protostars.

\begin{figure}
  \begin{center}
  \end{center}
  \caption{The same as Figure \ref{fig:figure19} 
but close-up for OMC-3.}\label{fig:figure20}
\end{figure}

\begin{figure}
  \begin{center}
  \end{center}
  \caption{The same as Figure \ref{fig:figure20} 
but with the positions of X-ray protostars 
\citep{tsu01}.}\label{fig:figure21}
\end{figure}

\begin{figure}
  \begin{center}
  \end{center}
  \caption{The Stamp map of the HC$_3$N profile around the 
X-ray protostars
observed by \citet{tsu01}.  Spectra are 4-channel (148 kHz) binned.
The X-ray protostars TKH8 and TKH10 are 
shown \citep{tsu01}}\label{fig:figure22}
\end{figure}

CCS was not detected in the OMC-2/3 region in either transitions.
We observed CCS over the same region as HC$_3$N (Figure \ref{fig:figure19}).
We also carried out a strip scan along 
Dec. (J2000) = $-$5$\degree$ 58$\arcmin$ 07$\arcsec$, 
over the right ascension range R. A. (J2000) = 5$^h$ 35$^m$ 3$^s$.8$-$33$^s$.2
passing through the southern part of the ``$\int$-shaped filament''
but have not detected in either transitions of CCS.
The 3$\sigma$ upper limit is 0.10 K in 45.4-GHz CCS 
and 0.25 K in 81.5-GHz CCS.
This means that the Orion A cloud is not very young,
which is consistent with a scenario that this cloud
was being compressed from the north
by the Orion superbubble, whose
first OB stars formed 1$-$2 $\times$ 10$^7$ yrs ago.
(\cite{bal87} and references therein).
There are other important observations showing the difference
between the Orion A cloud and the dark cloud TMC-1.
The neutral carbon atom C$^0$ and the CO isotopomer
show very similar distribution in Orion A and B clouds 
\citep{ike99,ike02}.
On the other hand, C$^0$ and 
the CO isotopomer
show quite different distribution in TMC-1.
The C$^0$ cloud is located on the south-east side
of the CO cloud containing TMC-1,
suggesting a chemical evolution from younger C$^0$ cloud
to CO cloud \citep{mae99}.
This difference will be due to a fact that
the Orion A cloud is rather older than TMC-1.
The C$^0$ gas in Taurus represents younger part of interstelar
cloud, while Orion clouds are more evolved without having
younger cloud.
The origin of C$^0$ in Orion is the photodissociation
region inside clumpy molecular clouds penetrated by the UV radiation.

The H$^{13}$ CO$^+$ core AC9 (\cite{aso00}, R. A. (J2000) = 
5$^h$ 35$^m$ 20$^s$.5, 
Dec. (J2000) = $-$5$\degree$ 5$\arcmin$ 13$\arcsec$) is 
less prominent 
in N$_2$H$^+$ and dust continuum.
We wonder whether this is a young core. 
Core AC9 is not prominent in HC$_3$N and is not observed in CCS.
There is no evidence that this core is young.
According to \citet{aso00} and \citet{wil03},
AC9 is located at or near the lobe of the intense outflow
from MMS9. We suspect that the H$^{13}$ CO$^+$ enhancement 
is due to
the effect of the intense outflow.

\section{Summary} 

We observed N$_2$H$^+$, HC$_3$N, and CCS in the Orion A cloud.
The N$_2$H$^+$ distribution was found to be very similar
to that of the dust continuum except for the central part
of the Orion Nebula.
The N-bearing molecules, N$_2$H$^+$ and NH$_3$ seem to be 
more intense 
in OMC-2, 
compared with the H$^{13}$CO$^+$ and CS distribution.
This suggests that OMC-2 has higher abundance of 
N-bearing molecules or higher filling factor of the quiescent gas.
We identified 34 molecular cloud cores on the basis of
N$_2$H$^+$ data.
The N$_2$H$^+$ excitation temperature in the Orion A cloud
is 1.6 times as high as that in the Taurus cores.
The excitation temperature decreases toward the south, suggesting
the core gas density or N$_2$H$^+$ abundance decreases 
toward the south.
The Orion cores have 
2.6, 3.2, 4.0 and 32 times larger core size,
linewidth, H$_2$ column density, and core mass than
Taurus cores, respectively.
The N$_2$H$^+$ linewidth shows variation along the ``$\int$-shaped 
filament.''
The N$_2$H$^+$ linewidth is large 
(1.1$-$2.1 km s$^{-1}$) over the Orion Nebula region,
moderate (0.5$-$1.3 km s$^{-1}$) in the OMC-2/3 region, 
and
smaller (0.3$-$1.1 km s$^{-1}$) in the south of Orion Nebula.
On the other hand, the gas kinetic temperature of the quiescent
cores observed in N$_2$H$^+$ is rather constant ($\sim$ 20 K).
The distribution of the HC$_3$N emission is globally similar
to that of N$_2$H$^+$.
We have not detected CCS anywhere in the OMC-2/3 region.
We found a peculiar starless core, AC9, which is intense in 
H$^{13}$CO$^+$,
but is not prominent in the dust continuum or N$_2$H$^+$.
The reason of H$^{13}$CO$^+$ enhancement could be
due to the effect of the lobe of the intense protostellar outflow
from the adjacent protostar MMS9.
We found a displacement between the N$_2$H$^+$ and HC$_3$N 
distribution
in OMC-3, which has a chain of Class 0 and/or Class I 
protostars
(and their candidates), including X-ray emitting protostars.
This displacement is likely to represent either the chemical evolution or
effect of protostellar outflows.

\bigskip

K. T. thanks Doug Johnstone for providing the FITS file
of the dust continuum emission.  The authors thank the 
referee, Dr. Edwin Bergin, for his
helpful comments.


\begin{longtable}{cccccccccccccl}
  \caption{N$_2$H$^+$ Core Catalog (1)}\label{tab:LTsample}
  \hline              
No. &  R. & A.  & (J2000) &  & Dec.  & (J2000)& $V_{LSR}$ & $R$ 
& $T_A^*$ & $T_{ex}$ & $\tau_{TOT}$ & $\Delta v$ & comment \\
\hline
 & h & m & s & $\degree$ & $\arcmin$ & $\arcsec$ & km s$^{-1}$  
& $\arcsec$ & K & K &  & km s$^{-1}$ &  \\
\endfirsthead
\hline
\endhead
  \hline
\endfoot
  \hline
\endlastfoot
  \hline
1 & 5 & 35 & 6.2 & -4 & 54 & 24 & 11.13  & 24  &  $\sim$ 1.3 
&  ... & ... &  0.65$\pm$0.05 & \\ 
2 & 5 & 35 & 6.1 & -4 & 56 & 7 & 11.19  & 43  & 1.34$\pm$0.33  
& 7.5$\pm$1.2  & 3.5$\pm$1.4  & 0.62$\pm$0.03 & \\ 
3 & 5 & 35 & 29.4 & -4 & 58 & 31 & 12.32  & 41  & 0.84$\pm$0.16  
& 5.6$\pm$0.5  & 3.8$\pm$1.2  & 1.31$\pm$0.10 & \\ 
4 & 5 & 35 & 19.8 & -5 & 0 & 53 & 11.29  & 42  & 2.60$\pm$0.11  
& 9.3$\pm$0.4  & 6.6$\pm$0.8  & 0.57$\pm$0.01 & \\ 
5 & 5 & 35 & 26.8 & -5 & 1 & 13 & 11.55  & 29  & 1.09$\pm$0.34  
& 7.5$\pm$1.5  & 2.5$\pm$1.1  & 0.79$\pm$0.05 & \\ 
6 & 5 & 35 & 25.4 & -5 & 2 & 36 & 11.13  & 49  & 1.57$\pm$0.10  
& 6.6$\pm$0.3  & 7.4$\pm$1.4  & 0.51$\pm$0.02 & \\ 
7 & 5 & 35 & 26.7 & -5 & 5 & 0 & 11.59  & 45  & 2.83$\pm$0.13  
& 10.0$\pm$0.4  & 6.2$\pm$0.7  & 0.47$\pm$0.01 & \\ 
8 & 5 & 35 & 32.3 & -5 & 6 & 2 & 11.84  & 24  & $\sim$ 1.5 
& ... & ...  & 0.74$\pm$0.04 & \\ 
9 & 5 & 35 & 23.9 & -5 & 7 & 25 & 11.88  & 48  & 1.98$\pm$0.29  
& 9.6$\pm$1.0  & 3.4$\pm$0.8  & 0.86$\pm$0.04 & \\ 
10 & 5 & 35 & 26.7 & -5 & 10 & 9 & 11.28  & 47  & 3.06$\pm$0.36  
& 20.3$\pm$1.9  & 1.7$\pm$0.2  & 1.23$\pm$0.02 & \\ 
11 & 5 & 35 & 22.6 & -5 & 10 & 9 & 11.67  & 26  & 2.41$\pm$0.43  
& 12.4$\pm$1.8  & 2.7$\pm$0.7  & 0.62$\pm$0.02 & \\ 
12 & 5 & 35 & 22.7 & -5 & 12 & 32 & 10.97  & 51  & 4.00$\pm$0.34  
& 18.6$\pm$1.3  & 2.7$\pm$0.3  & 0.93$\pm$0.02 & \\ 
13 & 5 & 35 & 21.2 & -5 & 14 & 36 & 10.82  & 46  & 2.31$\pm$0.21  
& 9.1$\pm$0.7  & 4.4$\pm$0.8  & 0.62$\pm$0.02 & \\ 
14 & 5 & 35 & 8.8 & -5 & 18 & 41 & 9.00  & 41  & 1.47$\pm$0.30  
& 7.9$\pm$1.1  & 3.5$\pm$1.2  & 0.62$\pm$0.03 & \\ 
15 & 5 & 35 & 15.8 & -5 & 19 & 26 & 9.94  & 55  & 3.51$\pm$0.15  
& 16.4$\pm$0.5  & 2.8$\pm$0.2  & 1.31$\pm$0.02 & \\ 
16 & 5 & 35 & 8.9 & -5 & 20 & 22 & 8.63  & 52  & $\sim$ 1.2 & ... 
& ...  & 2.05$\pm$0.09 & \\ 
17 & 5 & 35 & 10.3 & -5 & 21 & 25 & 8.11  & 36  & $\sim$ 1.7 & ... 
& ...  & 1.08$\pm$0.05 & \\ 
18 & 5 & 35 & 6.1 & -5 & 22 & 46 & 7.45  & 28  & $\sim$ 0.7 & ... 
& ...  & 2.06$\pm$0.14 & \\ 
19 & 5 & 35 & 12.9 & -5 & 24 & 10 & 6.58  & 36  & $\sim$ 1.0 & ... 
& ...  & 2.12$\pm$0.07 & \\ 
20 & 5 & 35 & 4.7 & -5 & 24 & 13 & 8.58  & 21  & $\sim$ 0.8 & ... 
& ...  & 1.92$\pm$0.10 & \\ 
21 & 5 & 35 & 15.7 & -5 & 25 & 54 & 8.23  & 30  & 1.32$\pm$0.42  
& 10.8$\pm$2.4  & 1.6$\pm$0.6  & 1.15$\pm$0.06 & \\ 
22 & 5 & 35 & 14.3 & -5 & 26 & 56 & 8.79  & 36  & $\sim$ 0.9 & ... 
& ...  & 1.72$\pm$0.07  & skew \\
23 & 5 & 35 & 2.0 & -5 & 36 & 10 & 7.27  & 57  & 1.52$\pm$0.03  
& 6.1$\pm$0.2  & 11.6$\pm$1.6  & 0.32$\pm$0.01 & \\ 
24 & 5 & 35 & 4.8 & -5 & 37 & 32 & 8.79  & 41  & 1.75$\pm$0.13  
& 7.3$\pm$0.4  & 6.2$\pm$1.2  & 0.72$\pm$0.03  & double peak \\
25 & 5 & 34 & 56.6 & -5 & 41 & 39 & 3.66  & 42  & 0.79$\pm$0.14 
& ... & ...  & 0.48$\pm$0.03 & \\ 
26 & 5 & 34 & 57.7 & -5 & 43 & 41 & 7.15  & 30  & 1.13$\pm$1.02  
& 9.8$\pm$6.5  & 1.5$\pm$1.7  & 0.41$\pm$0.03 & \\ 
27 & 5 & 34 & 56.3 & -5 & 46 & 5 & 5.52  & 36  & 0.68$\pm$0.16  
& 5.4$\pm$0.6  & 3.1$\pm$1.1  & 1.11$\pm$0.08 & \\ 
28 & 5 & 35 & 8.8 & -5 & 51 & 57 & 6.99  & 40  & $\sim$ 1.1 & ... 
& ...  & 0.38$\pm$0.05 & \\ 
29 & 5 & 35 & 0.7 & -5 & 55 & 40 & 8.03  & 30  & 0.89$\pm$0.46  
& 6.1$\pm$1.9  & 3.2$\pm$2.5  & 0.49$\pm$0.05 & \\ 
30 & 5 & 35 & 9.0 & -5 & 55 & 41 & 7.47  & 26  & 1.39$\pm$0.57  
& 9.2$\pm$2.7  & 2.3$\pm$1.3  & 0.64$\pm$0.04 & \\ 
31 & 5 & 35 & 12.8 & -5 & 58 & 6 & 7.62  & 41  & 0.67$\pm$0.62 
& ... & ...  & 0.83$\pm$0.09 & \\ 
32 & 5 & 35 & 28.1 & -6 & 0 & 9 & 7.29  & 78  & 0.57$\pm$0.08  
& 4.2$\pm$0.4  & 8.2$\pm$4.2  & 0.35$\pm$0.04 & \\
33 & 5 & 36 & 12.4 & -6 & 10 & 44 & 8.16  & 28  & 0.70$\pm$0.14  
& 4.7$\pm$0.5  & 5.6$\pm$2.5  & 0.62$\pm$0.06 & \\
34 & 5 & 36 & 24.7 & -6 & 14 & 11 & 8.19  & 36  & $\sim$ 0.5 
& ... & ...  & 0.92$\pm$0.16 & \\
  \hline
ave &  &  &  &  &  &  &  & 39$\pm$12  & 1.49$\pm$0.87  & 9.2$\pm$4.2 
& 4.3$\pm$2.5  & 0.92$\pm$0.52 &  \\
\end{longtable}

\begin{longtable}{ccccccccl}
  \caption{N$_2$H$^+$ Core Catalog (2)}\label{tab:LTsample}
  \hline              
No. & $R$ & $N$(N$_2$H$^+$) & $N$(H$_2$) & $n$(H$_2$) & $M$ 
& Paper I & Aso & Object \\
\hline
 & pc & cm$^{-2}$ & cm$^{-2}$ & cm$^{-3}$ & $M_{\odot}$ & & & \\
\endfirsthead
\endhead
  \hline
\endfoot
  \hline
\endlastfoot
  \hline
1 & 0.053  & ...     & ...     & ...     & ...   &   &     
&          \\ 
2 & 0.094  & 1.3E+13 & 4.2E+22 & 7.3E+04 & 25.9  &   &     
&          \\ 
3 & 0.089  & 2.5E+13 & 8.2E+22 & 1.5E+05 & 44.6  & 2 &     
& CSO3     \\
4 & 0.091  & 2.6E+13 & 8.8E+22 & 1.6E+05 & 50.0  & 3 & 3   
& H,MMS2-4,CSO5-6,VLA1 \\
5 & 0.064  & 1.2E+13 & 4.0E+22 & 1.0E+05 & 11.2  & 4 & 5   
&          \\ 
6 & 0.107  & 2.1E+13 & 6.9E+22 & 1.0E+05 & 54.4  & 5 & 7,8 
& H,MMS7   \\
7 & 0.098  & 2.2E+13 & 7.3E+22 & 1.2E+05 & 48.0  & 6 & 10  
& H,MMS8-9 \\
8 & 0.053  & ...     & ...     & ...     & ...   & 7 & 12  
& O,MMS10  \\
9 & 0.106  & 2.1E+13 & 7.1E+22 & 1.1E+05 & 54.6  & 8 & 14  
& O,OMC-2 FIR1a-c \\
10 & 0.102  & 2.8E+13 & 9.2E+22 & 1.5E+05 & 66.2 & 11 & 17 
& OMC-2 FIR4 \\
11 & 0.056  & 1.5E+13 & 5.0E+22 & 1.4E+05 & 10.7 &    & 16 
&            \\ 
12 & 0.111  & 3.1E+13 & 1.0E+23 & 1.5E+05 & 89.5 & 13 &    
&            \\ 
13 & 0.101  & 1.9E+13 & 6.4E+22 & 1.0E+05 & 44.9 & 16 &    
&            \\ 
14 & 0.089  & 1.4E+13 & 4.5E+22 & 8.2E+04 & 24.5 & 19 &    
&            \\ 
15 & 0.120  & 4.1E+13 & 1.4E+23 & 1.9E+05 & 135.9& 21 &    
&            \\ 
16 & 0.113  & ...     & ...     & ...     & ...  &    &    
&            \\ 
17 & 0.080  & ...     & ...     & ...     & ...  &    &    
&            \\ 
18 & 0.061  & ...     & ...     & ...     & ...  &    &    
&            \\ 
19 & 0.078  & ...     & ...     & ...     & ...  & 31 &    
& Ori-S \\
20 & 0.047  & ...     & ...     & ...     & ...   & 33 &  
&  \\ 
21 & 0.066  & 1.4E+13 & 4.8E+22 & 1.2E+05 & 14.6  & 36 &  
& \\ 
22 & 0.080  & ...     & ...     & ...     & ...   & 37 &  
& \\ 
23 & 0.124  & 2.0E+13 & 6.5E+22 & 8.5E+04 & 68.6  & 46 &  
& \\ 
24 & 0.089  & 2.6E+13 & 8.8E+22 & 1.6E+05 & 47.8  & 47 &  
&  \\ 
25 & 0.091  & ...     & ...     & ...     & ...   &    &  
& \\ 
26 & 0.066  & ...     & ...     & ...     & ...   & 51 &  
&  \\ 
27 & 0.080  & 1.6E+13 & 5.4E+22 & 1.1E+05 & 23.7  & 51 &  
& \\ 
28 & 0.087  & ...     & ...     & ...     & ...   & 54 &  
& \\ 
29 & 0.066  & ...     & ...     & ...     & ...   &    &  
& \\ 
30 & 0.056  & 1.0E+13 & 3.4E+22 & 9.7E+04 & 7.3   & 56 &  
& \\ 
31 & 0.089  & ...     & ...     & ...     & ...   & 57 &  
& \\ 
32 & 0.170  & 1.2E+13 & 4.0E+22 & 3.8E+04 & 78.0  &    &  
& \\ 
33 & 0.061  & 1.5E+13 & 5.1E+22 & 1.3E+05 & 13.1  & 61 &  
& \\ 
34 & 0.078  & ...     & ...     & ...     & ...   & 64 &  
& \\ 
  \hline
ave & 0.086$\pm$0.025  & $2.0\pm0.8\times10^{13}$ 
& $6.7\pm2.6\times10^{22}$ & $1.2\pm0.4\times10^5$ 
& 45.7$\pm$32.0  & & & \\ 
\end{longtable}


\end{document}